\documentclass{article}
\usepackage{ijcai26}

\usepackage{times}
\usepackage{soul}
\usepackage{url}
\usepackage[hidelinks]{hyperref}
\usepackage[utf8]{inputenc}
\usepackage[small]{caption}
\usepackage{graphicx}
\usepackage{amsmath}
\usepackage{amsthm}
\usepackage{booktabs}
\usepackage{algorithm}
\usepackage[switch]{lineno}

\usepackage{algpseudocode}
\usepackage{amsmath}
\usepackage{multirow}
\usepackage{amsfonts}
\usepackage[flushleft]{threeparttable}
	
\usepackage{graphicx}
\usepackage{subcaption}
\usepackage{xcolor}
\usepackage{pifont}
\usepackage{enumitem}
\usepackage{adjustbox}

\title{Light-FMP: Lightweight Feature and Model Pruning for Enhanced Deep Recommender Systems}

\author{
Nghia Bui$^1$
\and
Yue Ning$^2$\and
Lijing Wang$^1$\\
\affiliations
$^1$New Jersey Institute of Technology\\
$^2$Stevens Institute of Technology\\
\emails
\{ntd23, lw29\}@njit.edu,
yning5@stevens.edu
}

\begin{document}

\maketitle

\begin{abstract}
Deep recommender systems (DRS) often face challenges in balancing computational efficiency and model accuracy, especially when handling high-dimensional input features. 
Existing methods either focus on improving accuracy while neglecting training efficiency or prioritize efficiency at the cost of suboptimal accuracy across tasks. 
We propose Light-FMP: Lightweight Feature and Model Pruning for Enhanced DRS, a lightweight framework that
addresses the challenges through three key phases: \textit{pretraining}, \textit{pruning}, and \textit{continued training}. 
Using a hard concrete distribution, a masking layer is efficiently pretrained on a small data subset to identify important features. The model and features are then pruned, and training continues on the remaining dataset with domain-adapted parameters.
Experiments on benchmark datasets from real-world recommender systems demonstrate that Light-FMP outperforms existing methods in both efficiency and accuracy while maintaining scalability and robustness.
\end{abstract}

\section{Introduction}
\label{sec:introduction}
Recommender systems play an essential role in improving user experience and driving engagement on various digital platforms, including e-commerce \cite{linden2003amazon,schafer1999recommender,karthik2021fuzzy}, streaming services \cite{gomez2015netflix,covington2016deep,ko2022survey}, and social media \cite{jannach2019measuring,ricci2010introduction,kang2022ai,li2023survey}.  
Recommendation prediction estimates the likelihood that a user will interact with a recommended item (e.g., product or movie), based on features such as past interactions, preferences, and personal information \cite{zhong2019interpretable,liu2019pairwise,roy2022systematic,chen2023bias,li2022fairness}. These features serve as inputs to the prediction models. 

With advances in deep learning, deep learning-based recommender systems (DRS) have demonstrated remarkable accuracy in capturing user preferences, leveraging their powerful feature representation and inference capabilities~\cite{zhang2019deep,mu2018survey,da2020recommendation,anand2024survey,cai2024distributed}.  
However, DRS face significant challenges in computational efficiency, particularly when processing high-dimensional input features \cite{cheng2016wide,cheng2020differentiable,qu2023continuous} which range from dozens to thousands in small-scale systems and can exceed millions (e.g., Amazon~\cite{mcauley2015amazon}) in large-scale systems. The high-dimensional features often contain redundant or irrelevant information, leading to increased training costs, slower convergence, and overfitting \cite{rendle2010factorization}. Feature selection, the process of selecting a subset of informative features, has thus become essential for mitigating these issues and ensuring that only the most relevant features are used in model training~\cite{rong2019feature,wang2022autofield,lin2022adafs,lyu2023optimizing,wang2023single,lyu2022optembed,lee2023mvfs,jia2024erase}. 

\begin{figure*}[t]
\centering
\includegraphics[width=0.9\textwidth]{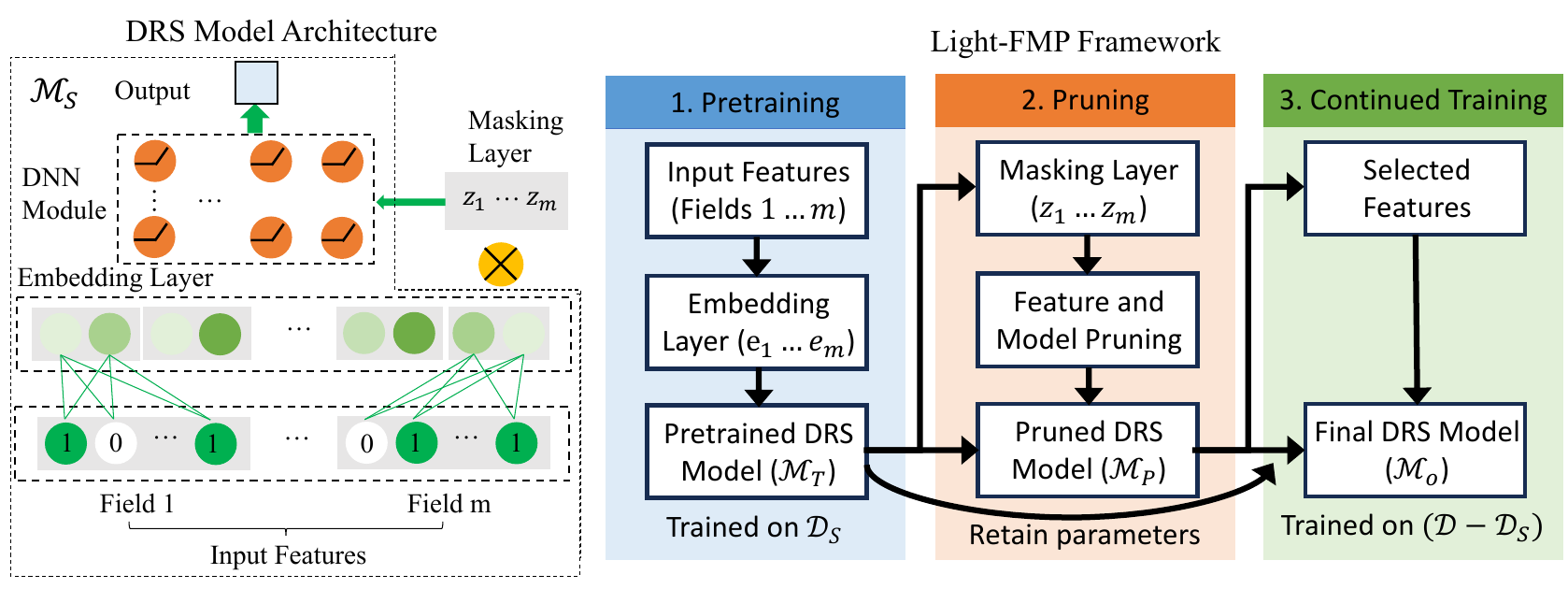}
  \caption{Overview of Light-FMP. The framework consists of three phases. Pretraining: A DRS model \(\mathcal{M}_S\) (the basic architecture is shown on the left) is augmented with a masking layer and trained on a small subset \(\mathcal{D}_S\), yielding the pretrained model \(\mathcal{M}_T\) and a learned mask. Pruning: The mask prunes both feature embeddings and \(\mathcal{M}_T\), producing selected features and model \(\mathcal{M}_P\). Continued training: \(\mathcal{M}_P\) is trained on the remaining dataset \(\mathcal{D} - \mathcal{D}_S\) using only pruned features, producing the final model \(\mathcal{M}_O\). }
  \label{fig:framework}
\end{figure*}

Existing feature selection methods for recommender systems can be broadly summarized into two classes: \textbf{shallow methods} and \textbf{deep methods}. \textit{Shallow methods} typically use statistical algorithms to assign feature importance score to each feature field, including traditional techniques such as Lasso\cite{tibshirani1996regression}, RandomForest \cite{leo2001random}, Gradient-boosted decision tree (GBDT) \cite{friedman2001greedy} and XGBoost \cite{chen2016xgboost}. These methods serve as static, task-agnostic preprocessing steps. While effective for simpler tasks, they struggle with high-dimensional and sparse data, leading to suboptimal performance in modern, dynamic recommender systems \cite{jia2024erase}.
\textit{Deep methods} leverage deep learning architectures or optimization strategies to dynamically select features through gate-based selection schema, typically following either single-stage or two-stage training paradigms. 
\textit{Single-stage methods}, such as AdaFS~\cite{lin2022adafs}, LPFS~\cite{guo2022lpfs}, and MvFS~\cite{lee2023mvfs}, integrate feature selection into model training. 
\textit{Two-stage methods}, including AutoField~\cite{wang2022autofield}, OptFS~\cite{lyu2023optimizing}, OptEm~\cite{lyu2022optembed}, and SFS~\cite{wang2023single}, separate feature selection and model training in two phases. These methods focus on improving accuracy by employing sophisticated feature selection mechanisms, but often neglect training and/or inference efficiency, making them unsuitable for large-scale applications. Furthermore, existing two-stage methods typically discard pretrained parameters after the first stage, missing the opportunity to improve accuracy by initializing the second stage with domain-adapted parameters, which is an oversight our work addresses.

To address these challenges, we propose \textbf{Light-FMP}: \textbf{Light}weight \textbf{F}eature and \textbf{M}odel \textbf{P}runing for Enhanced DRS, a novel three-phase framework to enhance DRS efficiency and accuracy through unified feature-model pruning. 
In \textit{pretraining}, a deep recommendation model is augmented with a masking layer, which learns feature importance via hard concrete distribution~\cite{louizos2017learning} (enable stable and efficient gradient-based optimization) on a small data subset, resulting in highly efficient pretraining. The \textit{pruning} phase uses this mask to compress both features and the model. In the \textit{continued training} phase, the pruned model initialized with domain-adapted parameters is trained on the full dataset using the reduced feature set, enabling comprehensive learning with enhanced efficiency. 
Compared to existing deep methods, Light-FMP not only enhances accuracy by selecting important input features but also greatly reduces computational overhead through efficient pretraining and domain-adapted parameter transfer. 

\noindent\textbf{Major contributions}: 
(1) We propose Light-FMP, a lightweight framework comprising pretraining, pruning, and continued training. This approach dynamically selects important features and trains a domain-adapted pretrained model through a light pretraining process, reducing computational overhead significantly while improving accuracy. 
(2) Unlike traditional two-stage methods, Light-FMP retains domain-adapted parameters from the pretraining phase, leading to improved accuracy in the continued training phase, which is a strategy not explored in prior works.
(3) We evaluate Light-FMP on benchmark datasets, demonstrating its superiority over existing shallow and deep feature selection methods in both computational efficiency and model accuracy. 
(4) The framework can be adapted to various backbone DRS models, making it broadly applicable across diverse domains.


    
    
    

\section{The Proposed Framework}
\label{sec:method}
This section details the Light-FMP framework. We begin with an overview of its key components and three-phase training process. We then describe the base DRS architecture, the masking layer, and the pretraining optimization. Next, we explain the pruning of both model and features, followed by the continued training phase, which leverages the pruned model and selected features for improved efficiency and accuracy.

\subsection{Framework Overview}\label{subsec:overview}
Figure~\ref{fig:framework} illustrates the Light-FMP framework overview.
Given a training dataset \( \mathcal{D} \) and a backbone DRS model \(\mathcal{M}_S\):
\textbf{Pretraining:}
A masking layer is inserted after the feature embedding layer in \(\mathcal{M}_S\) and trained on a randomly selected small subset \(\mathcal{D}_S \subset \mathcal{D}\) using an efficient optimization technique to identify important features. Simultaneously, \(\mathcal{M}_S\) learns domain-adapted parameters. The resulting masking layer and parameters are saved as the pretrained model \( \mathcal{M}_T \), ensuring a strong initialization for continued training phase.
\textbf{Pruning:}
The learned masking layer is used to prune both the input features and the pretrained model \( \mathcal{M}_T \),
yielding a compact pruned model \(\mathcal{M}_P\) without the masking layer. This pruning process reduces system complexity while preserving representational capacity, making the model suitable for large-scale datasets.
\textbf{Continued Training: }
The pruned model \(\mathcal{M}_P\) initialized with domain-adapted parameters is fine-tuned on the remaining dataset \( (\mathcal{D} - \mathcal{D}_S) \) using the reduced feature set. 
Compared to training from scratch (as in existing deep methods), the domain-adapted initialization enables improved predictive performance. 
The final model \(\mathcal{M}_O\) is both compact and highly effective. 

\subsection{Basic Architecture of DRS Models}\label{subsec:DRS}
In this subsection, we present the basic architecture of DRS models. As shown in the left part of Figure~\ref{fig:framework}, a typical DRS model comprises two key components: the embedding layer and the deep neural network (DNN) module. In our work, the backbone DRS models used in the experiments adhere to this basic architecture. Our framework is built upon this foundational design and is adaptable to any backbone DRS model that follows it.


\noindent\textbf{Embedding Layer.}
In web-scale recommender systems, input features are high-dimensional, sparse, and grouped into fields, which can be either categorical (e.g., gender) or continuous (e.g., price). A categorical field is represented as a sparse binary vector $\mathbf{x}_m \in \mathbb{R}^{D_m}$, where $D_m$ is the number of its unique values. A continuous field is represented as a scalar value $x_m \in \mathbb{R}$ (i.e., $D_m=1$). 
Given an input feature \( \mathbf{x} \) of \( m \) feature fields, the full vector is the concatenation of the representations from all 
$m$ fields, denoted as $\mathbf{x} = [\mathbf{x}_1,\dots,\mathbf{x}_{k},x_{k+1},\dots,x_m]$, where the first $k$ fields are categorical and the remaining $m-k$ are continuous.
The embedding layer transforms the sparse input feature vector \( \mathbf{x} \) into dense, continuous vectors of lower dimensions through projection \( \mathbf{e}_m = A_m \mathbf{x}_m\) where \(A_m \in \mathbb{R}^{D\times D_m}\) is a learnable matrix with \( D \) as the embedding dimension, capturing meaningful patterns and relationships between features. The final representation is the concatenation of all field embeddings:
\begin{equation}
\label{equ:embedding}
\mathbf{e} = [\mathbf{e}_1, \mathbf{e}_2, \dots, \mathbf{e}_{m-1}, \mathbf{e}_m] \in \mathbb{R}^{mD}
\end{equation}

\noindent\textbf{DNN Module.}
Feature interactions in DRS are modeled through a DNN module typically a Multi-Layer Perceptron (MLP) that processes dense embeddings $\mathbf{e}$. This MLP uses stacked fully connected layers with non-linear activations to capture both low- and high-order dependencies between feature fields. The output layer produces the prediction \(\hat{\mathbf{y}}\) by applying a task-specific activation function, e.g., softmax for classification. 


\noindent\textbf{Optimization.}
Given a training dataset \(\mathcal{D}=\{X, \mathbf{y}\}=\{\mathbf{x}^i, y^i\}_{i=1}^{n}\) of \( n \) training examples.
\(y^i\) depends on the specific task. For instance, in a Click-Through Rate (CTR) prediction problem, \(y^i\) is a binary label indicating whether a user clicked (\(y^i=1\)) or did not click (\(y^i=0\)) on a given item. 
The learning objective is to learn a model \( f_{\theta} \) parameterized by \( \theta \) that maps from input \( X \) to target \( \mathbf{y} \) by minimizing a loss function \(\mathcal{L}(\theta)\), typically a log-loss for classification tasks and a \(L_1\) loss or \(L_2\) loss for regression tasks.





\subsection{Pretraining}\label{subsec:pretraining}
\noindent\textbf{Masking Layer.}
Given a backbone DRS model \(\mathcal{M}_S\), we add a masking layer, denoted as a masking variable \(\mathbf{z} = [z_1,  \cdots,  z_{m}] \in \mathbb{R}^m\), after the embedding layer. Its primary function is to selectively mask or sparsify the features, allowing the model to focus on more relevant information and reduce computational complexity.
During the pretraining, \textit{in forward pass}, the dense feature embedding \( \mathbf{e}\) will be masked by \(\mathbf{z}\):

\begin{equation}
\label{equ:masking}
\mathbf{e}' = \mathbf{e} \odot \mathbf{z}  \in \mathbb{R}^{mD}
\end{equation}

\noindent where \( \odot \) is the element-wise product. 
The masked embedding  \( \mathbf{e}'\) will be passed to the DNN module to perform the regular feature interaction and transformation steps as described in Section~\ref{subsec:DRS}.
\textit{In the backpropagation pass}, \(\mathbf{z}\) is optimized together with all other model parameters \(\theta\). Each element in the masking variable is learned as a binary value, serving as a gate that determines whether the corresponding field embedding should be pruned or retained. 

\textit{Existing deep methods use gate-based feature selection relying on non-differentiable tricks (e.g., Gumbel-Max in AutoField, max-pooling in AdaFS), complex scheduling (OptFS), or regularization (OptFS, OptEm), which often lead to poor scalability, unstable convergence, and high training complexity. Inspired by \cite{xia2023sheared}, we instead adopt the differentiable hard concrete distribution~\cite{louizos2017learning} for stable, efficient gradient-based optimization and fast pretraining.}
The specific learning process of \(\mathbf{z}\) in our method is shown in Algorithm~\ref{alg:learning_z}. 
The masking variable \(\mathbf{z}\) is parameterized by \( \alpha \):
\begin{equation}
\begin{aligned}
\label{equ:learning_z}
\alpha &\sim \mathcal{N}(\mu, \sigma ^2) \in \mathbb{R}^{m}\\
u &\sim \mathcal{U}(0, 1) \in \mathbb{R}^{m}\\
\mathbf{s} &= \text{sigmoid}\left( \frac{1}{\beta} \left( \log \frac{u}{1 - u} + \log \alpha \right) \right) \in \mathbb{R}^{m}\\
\bar{\mathbf{s}} &= \mathbf{s} (\zeta - \gamma) + \gamma \in \mathbb{R}^{m}\\
\mathbf{z} &= \min(1, \max(0, \bar{\mathbf{s}})) \in \mathbb{R}^{m}\\
\end{aligned}
\end{equation}

\noindent Here, \(\mathcal{N}\) is the normal distribution, \(\mathcal{U}\) is the uniform distribution, \(\beta\) is a temperature parameter, and \(s\) is a relaxed binary mask following a hard concrete distribution bounded by \(\gamma < 0\) and \(\zeta > 1\). Following prior works~\cite{wang2020structured,xia2023sheared}, this distribution constrains \(s\) to \([\gamma, \zeta]\), concentrating probability mass near 0 and 1 after rectification, which promotes binarization. While the hard concrete distribution drives the mask variable \(\mathbf{z}\) toward binary values, a final thresholding step is applied during pruning to ensure strictly binary masks in practice.
In our experiments, we set \(\mu = 0.5\), \(\sigma^2 = 0.0625\), \(\beta = 0.83\), \(\zeta=1.1\), and \(\gamma=-0.1\), from literature experiment in \cite{xia2023sheared}.
\noindent\textbf{Optimization.}
In our work, pruning is framed as a constrained optimization problem, where the model learns \(\mathbf{z}\) to identify an important feature subset that optimizes performance. 
The loss term for optimizing \(\mathbf{z}\) is defined as: 
\begin{equation}
\label{equ:loss_z}
\tilde{\mathcal{L}}(\lambda, \phi, \mathbf{z}) = \lambda (\frac{1}{m}\sum \mathbf{z} - \tau) + \phi (\frac{1}{m}\sum \mathbf{z} - \tau)^2    
\end{equation}

\noindent where \((\lambda, \phi)\) is a pair of Lagrange multipliers to impose constraints on the model \(\mathcal{M}_S\), \(\tau = \frac{\text{number of pruned fields}}{m}\) is the pruning ratio, controlling the proportion of embedding vectors that are pruned from the model. The total loss in pretraining phase is: 
\begin{equation}
\label{equ:loss_total}
\mathcal{L}_{\text{total}} (\theta, \lambda, \phi, \mathbf{z}) = \mathcal{L} (\theta, \mathbf{z}) + \tilde{\mathcal{L}}(\lambda, \phi, \mathbf{z})
\end{equation}

\begin{algorithm}[t]
\caption{Learning the Masking Variable \(\mathbf{z}\)}
\label{alg:learning_z}
\begin{algorithmic}[1]
\Require Model parameters \(\theta\), masking variable \(\mathbf{z}\), Lagrange multipliers \((\lambda, \phi)\), pruning ratio \(\tau\), temperature parameter \(\beta\), bounds \(\zeta, \gamma\), learning rate \(\eta\)
\State Initialize \(\alpha \sim \mathcal{N}(\mu, \sigma^2)\)
\Repeat
    \State Sample \(u \sim \mathcal{U}(0, 1)\)
    \State Compute relaxed mask \(\mathbf{s} \gets \text{Sigmoid}\left(\frac{1}{\beta} \left( \log \frac{u}{1 - u} + \log \alpha \right)\right)\)
    \State Adjust mask \(\bar{\mathbf{s}} \gets \mathbf{s} \cdot (\zeta - \gamma) + \gamma\)
    \State Compute binary mask \(\mathbf{z} \gets \min(1, \max(0, \bar{\mathbf{s}}))\)
    \State Compute model loss \(\mathcal{L}(\theta, \mathbf{z})\)
    \State Compute constraint loss \(\tilde{\mathcal{L}}(\lambda, \phi, \mathbf{z})\) based on Equation~\ref{equ:loss_z}
    \State Compute total loss \(\mathcal{L}_{\text{total}}\) based on Equation~\ref{equ:loss_total}
    \State Update model parameters \(\theta \gets \theta - \eta \nabla_\theta \mathcal{L}_{\text{total}}\)
    \State Update masking variable parameter \(\alpha \gets \alpha - \eta \nabla_\alpha \mathcal{L}_{\text{total}}\)
\Until{Convergence}
\end{algorithmic}
\end{algorithm}

\subsection{Pruning}\label{subsec:pruning}

Given the learned masking vector \( \mathbf{z} \in \mathbb{R}^m\) and the domain-adapted model \(\mathcal{M}_T\) from pretraining, the pruning phase removes input features and model parameters where \( z_m = 0 \). The input \( \mathbf{x} = [\mathbf{x}_1,\dots, \mathbf{x}_m] \) is pruned to retain only fields where \(z_m=1\). A new model \(\mathcal{M}_P\) is initialized with the same architecture as the base model \(\mathcal{M}_S\) and pruned by removing all embedding vectors and their associated connections where \( z_m = 0 \). The weights of \(\mathcal{M}_P\) are then set to the corresponding saved weights from \(\mathcal{M}_T\) , resulting in a compact, domain-adapted model without a masking layer. This weight transfer provides a strong initialization for continued training. The process reduces model complexity, lowers computational cost, and can improve final accuracy.

\subsection{Continued Training }\label{subsec:continued_training}

The continued training phase fine-tunes the pruned model $ \mathcal{M}_P $ to optimize performance. Because pretraining on the subset $ \mathcal{D}_S $ only established initial parameters and the masking variable $\mathbf{z}$, and pruning introduced structural changes, further adaptation is necessary. Consequently, $ \mathcal{M}_P $ is fully fine-tuned on the remaining dataset $ (\mathcal{D} - \mathcal{D}_S) $. 
Leveraging reduced complexity and domain-adapted parameters, this phase achieves better convergence and lower computational costs than existing two-stage methods.

\subsection{Model Inference}\label{subsec:predicting}
After pretraining, pruning, and continued training, we obtained a well-trained model \( \mathcal{M}_O \) for inference on new data. The input features \( \mathbf{x} \) are pruned according to the masking variable \( \mathbf{z} \). Unlike existing methods (e.g., AutoField, AdaFS, OptFS, MvFS, LPFS) which retain full model for inference, our framework produces a compact model \( \mathcal{M}_O \) with reduced size, enabling faster inference for real-time applications without sacrificing accuracy.

\subsection{Significance of the Proposed Framework}\label{subsec:significance}
\textbf{Light-FMP significantly improves training and inference efficiency} through carefully designed optimizations compared with existing deep methods.
\begin{itemize}[leftmargin=*]
    \item \textit{Super efficient pretraining}: Leverages the hard concrete distribution on a very small dataset subset for stable, low-overhead gradient-based optimization. This identifies important features and learns domain-adapted parameters for strong initialization.
    \item \textit{Accelerated continued training}: Unlike methods retaining the full model, we use a pruned model $ \mathcal{M}_P$ with reduced dimensions, lowering per-step computational costs.
    \item \textit{Reduced inference time}: We deploy a compact final model $ \mathcal{M}_O$, achieving faster inference without sacrificing performance, making it suitable for real-time systems.
\end{itemize}
Compared to training from scratch, \textbf{our method enhances predictive performance} by retaining informative features and leveraging pretrained domain-adapted parameters to improve generalization and accuracy. \textbf{Furthermore, our three-phase framework is modular and easy to execute.} Phases run independently or sequentially with minimal complexity, resulting in a compact, efficient, and accurate model for large-scale DRS.

\section{Experiment}
In this section, we describe our experimental settings and compare our approach with state-of-the-art (SOTA) baselines to demonstrate the effectiveness of LightFMP. 
Additional sensitivity analysis experiments reveal the consistency of the redundant features.

\subsection{Data \& Metrics}\label{subsec:data_metric}


In our experiments, we use datasets commonly used for benchmarking CTR prediction in recommender systems: \textbf{Criteo}~\cite{criteo2014challenge}, \textbf{Avazu}~\cite{zhu2021open}, and \textbf{MovieLens}~\cite{harper2015movielens}.
We randomly split the data into training, validation, testing, and pretraining subsets, preserving the original class distribution.
The statistics of these datasets are summarized in Table~\ref{tab:data_stat}. 
\begin{table}[h]
    \small
    \centering
    \begin{tabular}{lccccc}
        \toprule
        Dataset            & Total         & \#Train      & \#Val  & \#Test  &\# Pretrain       \\
        \midrule
        Criteo     & 45.8M         & 33.0M        & 8.3M         & 4.6M  & 2K           \\
        Avazu         & 40.4M         & 32.3M        & 4.0M         & 4.0M   & 2K          \\
        MovieLens         & 20.0M         & 16.0M        & 2.0M         & 2.0M   & 2K          \\
        \bottomrule
    \end{tabular}
    \label{tab:dataset_statistics}
    \begin{tablenotes}
      \small
      \item Due to missing details, exact data splits from various baseline methods are unreproducible, but we ensure reliable comparisons by using consistent splitting across all baseline implementations in our paper.
    \end{tablenotes}
    \caption{Datasets used in experiments.}
    \label{tab:data_stat}
\end{table}

\begin{table*}[t]
\centering
\begin{adjustbox}{max width=\textwidth}
\begin{tabular}{l|ccc|ccc|ccc}
\toprule
\textbf{Selection approach} & \multicolumn{3}{c}{\textbf{Criteo}} & \multicolumn{3}{c}{\textbf{Movielens}} & \multicolumn{3}{c}{\textbf{Avazu}} \\
\cmidrule(r){2-4} \cmidrule(r){5-7} \cmidrule(r){8-10}
& \textbf{AUC} $\uparrow$ & \textbf{Logloss} $\downarrow$ & \textbf{TT} $\downarrow$ & \textbf{AUC} $\uparrow$ & \textbf{Logloss} $\downarrow$ & \textbf{TT} $\downarrow$  & \textbf{AUC} $\uparrow$ & \textbf{Logloss} $\downarrow$ & \textbf{TT} $\downarrow$ \\
\midrule
No Selection & 0.7913& 0.4616& 1.0000 & 0.8045& 0.5357& 1.0000 & 0.7782& 0.3807& 1.0000\\
\cmidrule{1-10}
 Lasso (\textit{shallow}) & 0.7977& 0.4531& 0.6482& 0.7163& 0.6178& 0.5994& 0.7600& 0.3907& 0.6480\\
 PCA (shallow)& 0.7965& 0.4543& 0.7007& \underline{0.8062}& \underline{0.5334}& 0.7093& 0.7587&  0.3911&0.4742\\
 XGBoost (\textit{shallow})& 0.7746& 0.4721& 0.5866& 0.7979& 0.5426& 0.6849& 0.7650& 0.3878& 0.7420\\
 OptFS (\textit{deep})& \underline{0.8036}& \underline{0.4475}& 1.0761& 0.7176& 0.6166& 1.3974& 0.7808& 0.3795& 0.6466\\
 OptEm (\textit{deep})& \textbf{0.8043}& \textbf{0.4468}& 1.2672& 0.7171& 0.6168& 1.9803& 0.7807& 0.3800& 1.5600\\
 Autofield (\textit{deep})& 0.7996& 0.4516& 1.1896& 0.8051& 0.5352& 0.7615& \underline{0.7824}& 0.3788& 0.4291\\
 AdaFS (\textit{deep})& 0.7934& 0.4568& \textbf{0.3368}& 0.7185& 0.6166& \textbf{0.3189}& 0.7758& 0.3829& \textbf{0.3213}\\
 MvFS (\textit{deep}) & 0.7956& 0.4550& 1.0889& 0.7171& 0.6170& \underline{0.3329}& 0.7819& \underline{0.3787}& \underline{0.3365}\\
\cmidrule{1-10}
 \textbf{Light-FMP} (\textit{deep})& 0.8022& 0.4486& \underline{0.6084}& \textbf{0.8071}& \textbf{0.5324}& 0.7080& \textbf{0.7845}& \textbf{0.3777}& 0.6037\\
\bottomrule
\end{tabular}
\end{adjustbox}
\caption{Overall experimental results of feature selection using xDeepFM backbone. The pruning ratio for all methods except No Selection and MvFS is set to 0.5. Time values are represented as ratios relative to the original training time: 15,803 seconds for Criteo and 6,323 seconds for Movielens and 15,312 seconds for Avazu. Bold face indicates the best results of each column. The results are an average of 3 random runs. }
\label{tab:overall-xdeepfm}
\end{table*}

We utilize the most widely used evaluation metrics for CTR prediction: \textbf{AUC (Area Under the ROC Curve)} and \textbf{Logloss}. 
It should be emphasized that \textit{a higher AUC score and a lower Logloss at the 0.001 level signify a \textbf{substantial improvement}; otherwise a \textbf{marginal improvement} if less than 0.001} in CTR prediction tasks~\cite{guo2017deepfm,wang2022autofield,lin2022adafs}. 
Furthermore, we evaluate the efficiency of a model by its \textbf{Pretraining Time (PT)}, \textbf{Continued Training Time (CT)}, \textbf{Total Training Time (TT)}, and \textbf{Inference Time (IT)}.

\subsection{Baselines}\label{subsec:baseline}
To evaluate the effectiveness of Light-FMP in feature selection, we implement the following feature selection baselines.
\textbf{Shallow methods:} 
\textbf{Lasso} \cite{tibshirani1996regression} is a classic method for variable selection and regularization. 
\textbf{PCA}~\cite{pearson1901liii} is a feature decorrelation aims to reduce unimportant features.
 \textbf{XGBoost} \cite{chen2016xgboost} ranks features based on how much they improve the final model performance.  
\textbf{Deep methods:} 
\textbf{AutoField}  \cite{wang2022autofield} selects features at the field level using a controller network.  
\textbf{AdaFS} \cite{lin2022adafs} adaptively generates gates for feature fields at the field level.
\textbf{MvFS} \cite{lee2023mvfs} advances AdaFS by mitigating bias problem to major features. 
\textbf{OptFS} \cite{lyu2023optimizing} trains scalars and enforces sparsity constraints at the value level. 
\textbf{OptEm} \cite{lyu2022optembed} optimizes feature selection by dynamically adjusting criteria and minimizing computational costs.

\subsection{Backbone DRS Models}\label{subsec:backbone}
In our paper, we leverage five widely adopted DRS backbone models: \textbf{xDeepFM} \cite{lian2018xdeepfm}, \textbf{DeepFM} \cite{guo2017deepfm}, \textbf{Wide\&Deep} \cite{cheng2016wide}, \textbf{PNN} \cite{qu2016product}, and \textbf{DNN} \cite{lecun2015deep}. These models represent diverse architectures commonly employed in recommendation tasks, offering a comprehensive basis for evaluating the adaptability and effectiveness of our framework across different DRS models. 

\subsection{Implementation Details}\label{subsec:implemenation} 
Our implementation is based on the open-source code from OptFS~\cite{lyu2023optimizing}, using a learning rate tuned from ${1e-5, 1e-4, 3e-4}$
and Adam optimizer with $ \lambda = 1e-5 $ regularization for both linear and embedding layers. Embedding dimension $D$ is 10 for Criteo and Avazu, 64 for Movielens. The prune ratio $ \tau $ defaults to 0.5 and varies in sensitivity analysis. The fully connected network consists of layers with 400, 400, and 200 neurons. Results are averages of 3 random trials. 

\section{Result}

\subsection{Overall Performance}\label{subsec:performance}
This section evaluates Light-FMP against feature selection baselines. Table~\ref{tab:overall-xdeepfm} presents the performance of various feature selection methods integrated into the xDeepFM, evaluated on the Criteo, Movielens, and Avazu datasets in terms of accuracy (AUC and Logloss) and total training time (TT). 
For all two-stage methods (including shallow baselines, OptFS, OptEm, AutoField, and Light-FMP), training time includes both feature selection and retraining.
On Criteo, Light-FMP achieves competitive accuracy (0.8022 AUC, 0.4486 Logloss) compared to the best baselines i.e., OptEm (0.8043 AUC, 0.4468 Logloss) and OptFS (0.8036 AUC, 0.4475 Logloss), while significantly improving efficiency. Light-FMP reduces total training time to 60.84\% of the unpruned baseline, significantly outperforming OptEm (1.2672 TT) and OptFS (1.0761 TT). This pattern is even more pronounced on Movielens, where OptEm (1.3974 TT) and OptFS (1.9803 TT) require substantially more training time, yet deliver poor accuracy (approximately 0.72 AUC and 0.62 Logloss compared to the No Selection baseline of 0.8045 AUC and 0.5357 Logloss).
On the contrary, on Movielens and Avazu, Light-FMP achieves the best accuracy (0.8071 AUC and 0.5324 Logloss on Movielens; 0.7845 AUC and 0.3777 Logloss on Avazu) and significantly reduce total training time compared to No Selection baseline (70.80\% and 60.37\% reductions in TT). 
The biggest time reductions are achieved by AdaFS on Criteo (33.68\%), Movielens (31.89\%), and Avazu (32.13\%), nearly half the training time of our method. However, the accuracy is substantially lower than our method in both AUC and Logloss on three datasets. MvFS exhibits a similar trade-off, sacrificing accuracy for efficiency. 



AdaFS and MvFS achieve high efficiency via category-level pruning, compressing embeddings by removing low-importance categorical values. However, MvFS's training time varies widely due to dataset-dependent prunable category ratios, e.g., fewer are removed on Criteo (slower training), while more are pruned on Movielens and Avazu (faster training). AutoField similarly exhibits inconsistent efficiency because its data-driven feature masking leads to non-uniform pruning levels across datasets, e.g., mild pruning on Criteo increases training time by 20\%, whereas aggressive pruning on Avazu and Movielens cuts it to 42\% and 76\% of the baseline. These results highlight the sensitivity of both category-level and learned masking methods to dataset structure, in contrast to Light-FMP's more stable feature-level pruning.

Compared to two-stage deep methods (OptFS, OptEm, Autofield), Light-FMP achieves better or comparable accuracy while requiring consistently lower training time. Compared to single-stage deep methods (AdaFS, MvFS), Light-FMP significantly outperforms in accuracy. Although shallow methods show robust training time reduction across datasets, they exhibit large variation in accuracy performance, indicating their limited capacity to adapt to complex feature distributions.
\textit{Overall, Light-FMP consistently strikes the best accuracy-efficiency trade-off across different tasks, providing scalable and effective feature selection for DRS.}

\subsection{Computational Cost}\label{subsec:efficiency}
\begin{figure}[t]
\centering
    \begin{subfigure}[t]{0.48\columnwidth}
        \centering
        \includegraphics[width=\linewidth]{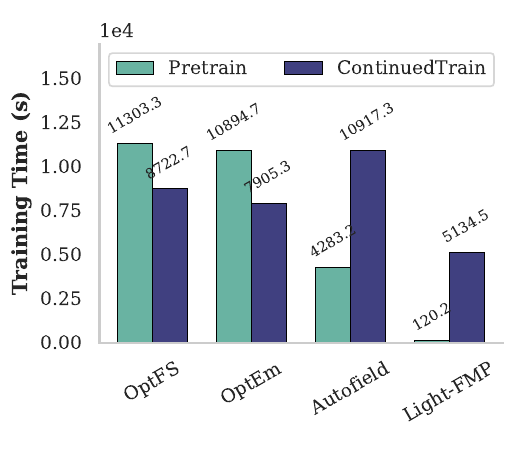} 
          \caption{Pretraining and continued training time (in seconds) of two-stage deep methods.}
        \label{fig:train_time}
    \end{subfigure}
    \hfill
    \begin{subfigure}[t]{0.48\columnwidth}
        \centering
        \includegraphics[width=\linewidth]{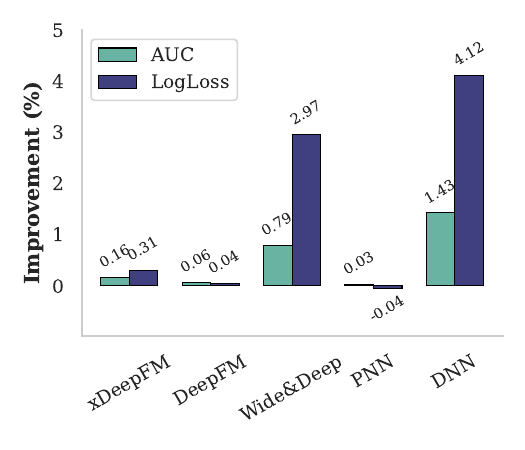} 
          \caption{Ablation study on domain-adapted parameter initialization.}
        \label{fig:init_ablation}
    \end{subfigure}
    \hfill
    \begin{subfigure}[t]{0.48\columnwidth}
        \centering
        \includegraphics[width=\linewidth]{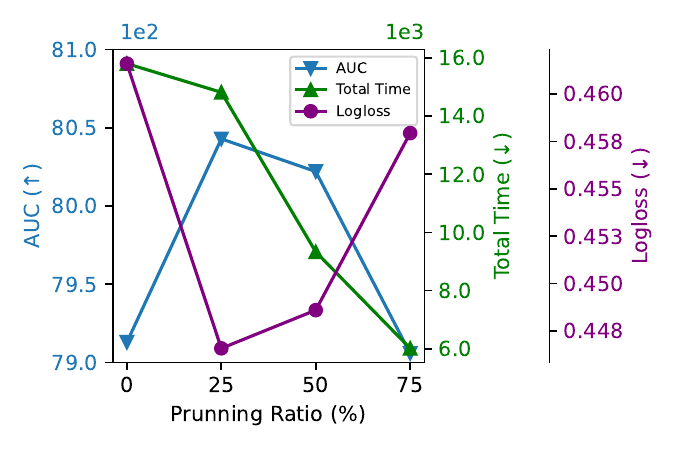}
        \caption{Sensitivity to pruning ratio: 0\%, 25\%, 50\%, 75\%.}
        \label{fig:pruning_ratio}
    \end{subfigure}
    \hfill
    \begin{subfigure}[t]{0.48\columnwidth}
        \centering
        \includegraphics[width=\linewidth]{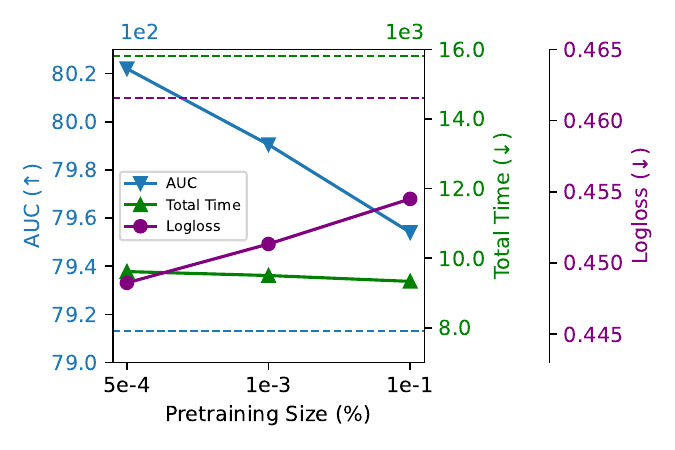}
        \caption{Sensitivity to pretraining size: 5e-4\%, 1e-3\%, 1e-1\%, dashed lines represent xDeepFM}
        \label{fig:pretrain_size}
    \end{subfigure}
    \caption{Experimental results on Criteo using xDeepFM. 
    }
    \label{fig:analysis}
\end{figure}

Table~\ref{tab:overall-xdeepfm} reports the overall training cost. The results show that Light-FMP achieves the highest total training efficiency compared to all two-stage deep baselines (OptFS, OptEm, Autofield).
To identify the primary source of this efficiency gain, we break down the pretraining and continued training times for two-stage deep methods on Criteo and present the results in Figure~\ref{fig:train_time}. Light-FMP requires only 120 seconds for pretraining, far lower than the second best method, Autofield (4283 seconds); Light-FMP also achieves the lowest continued training time at 5134.5 seconds, notably faster than the next best, OptEm, which requires 7905.3 seconds. These results highlight Light-FMP’s strong efficiency advantage of both stages.
A key reason for Light-FMP's highly efficient pretraining is that it operates on a very small subset of data (2K samples out of 45M on Criteo), unlike OptEm, and Autofield which require searching for optimal mask on the entire dataset. This efficiency extends to continued training, as Light-FMP uses the remaining subset for this phase, whereas other baselines train on the full set. 
The training efficiency gains become even more substantial with large-scale, real-world datasets and high-dimensional features.
 
In addition, unlike No Selection, OptFS, AutoField, and OptEm, whose large numbers of features lead to slow inference, Light-FMP maintains a compact model, enabling efficient inference while reducing storage requirements. This advantage becomes more pronounced in high-dimensional feature spaces.

\begin{table}[t]
\centering
\begin{tabular}{lccc}
\toprule
\multicolumn{1}{c}{\textbf{Backbone}} & \multicolumn{3}{c}{\textbf{Criteo}}\\

	& \textbf{AUC} $\uparrow$ 	& \textbf{Logloss} $\downarrow$ 	& \textbf{TT} $\downarrow$ 	\\

\midrule							
xDeepFM  	&0.7913&0.4616&4h23m23s\\
\quad + Light-FMP  	& 0.8022& 0.4486& 2h40m15s\\
Improvement (\%) &	1.37&2.81&39.15\\
\midrule
DeepFM  	&0.7891& 0.463&5h5m12s\\
\quad + Light-FMP  	& 0.7941&0.4557& 2h25m46s\\
Improvement (\%) &0.63&	1.58& 53.06\\
\midrule
Wide\&Deep	&0.7746& 0.4833&5h2m43s\\
\quad + Light-FMP  	&0.7888&0.462&  2h36m50s\\
Improvement (\%) &1.83&4.41&48.19\\
\midrule
PNN  	&0.78&0.4801&3h32m52s\\
\quad + Light-FMP 	&0.7952&0.4551&2h37m2s\\
Improvement (\%) &1.95&5.21&26.22\\
\midrule
DNN  	&0.7737&0.4838&4h52m26s\\
\quad + Light-FMP 	& 0.7847& 0.464& 2h25m36s\\
Improvement (\%) &1.42&4.09& 50.21\\
\bottomrule
\end{tabular}
\caption{Adaptability performance on Criteo. Improvement percentages reflect the gains achieved by applying our framework to the backbone models.  Bold face indicates the best results.}
\label{tab:adaptability_criteo}
\end{table}

\subsection{Adaptability Study}\label{subsec:adaptability}
In this section, we show that our Light-FMP is model-agnostic, allowing it to seamlessly integrate with various backbone architectures. We evaluated the adaptability of our framework using five popular DRS backbone models, including xDeepFM, DeepFM, Wide\&Deep, PNN, and DNN, on the Criteo dataset in terms of accuracy (AUC and Logloss), and efficiency (Total Training Time (TT)). The results are shown in Table~\ref{tab:adaptability_criteo}.
\textbf{For accuracy}, Light-FMP leads to a substantial improvement in AUC and Logloss across all backbone models, such as Logloss improves notably for PNN (5.21\%) and Wide\&Deep (4.41\%).  
\textbf{For efficiency}, Light-FMP demonstrates substantial reductions in training time, with training time improvements ranging from 26.22\% (PNN) to 53.06\% (DeepFM). 
Overall, these results highlight that our method not only enhances the accuracy of DRS but also significantly boosts computational efficiency, making it highly adaptable across different backbone models and datasets.

\subsection{Sensitivity Analysis}\label{subsec:sensitivty}
We investigate how the pruning ratio $\tau$ and the pretraining size affect the accuracy of our method.

\subsubsection{\textbf{Pruning Ratio \( \tau \)}}\label{subsubsec:pruning_ratio}
We explore the effect of pruning ratio \( \tau \) on model performance by running Light-FMP with the same settings while varying \(\tau\) as 0 (no pruning), 25\%, 50\%, and 75\% (Figure~\ref{fig:pruning_ratio}).
Accuracy improves when \( \tau \) increases from 0\% to 25\%, but declines with further pruning (50\% and 75\%). This indicates that moderate pruning ($<$ 25\%) can enhance generalization, while excessive removal ($>$ 50\%) likely discards important features and harms performance.
However, determining the optimal pruning ratio remains challenging, as it depends on the specific dataset, task, and model architecture. Adaptively adjusting the pruning ratio is a promising direction for future work.

Total training time consistently decreases as \( \tau \) increases, demonstrating clear efficiency gains from pruning.
The results suggest that removing redundant features reduces model complexity, mitigates overfitting, and focuses learning on informative signals, thereby improving generalization, especially under light pruning.

\subsubsection{\textbf{Pretrain Size}}\label{subsubsec:pretrain_size}
Figure~\ref{fig:pretrain_size} show the sensitivity results on Criteo using xDeepFM with varying pretrain sizes (0.0005\%, 0.001\%, 0.1\%). 
Notably, model performance in terms of AUC and Logloss consistently declines as pretrain size increases, yet remains above the No Selection baseline. This suggests that Light-FMP achieves performance gains even with minimal pretraining data. While a larger pretraining set can further improve training efficiency, it may compromise predictive performance. As detailed in the computational cost analysis, a larger pretraining subset leads to a smaller continued training set. This reduces total training time but can also lower final accuracy, as the compact model learns primarily during the continued training phase rather than pretraining.
Based on this insight, our experiments adopt a small fixed pretraining size (2K samples) across all three benchmark datasets, rather than a percentage-based approach. This configuration yields consistent improvements in both predictive accuracy and training efficiency. 

\subsubsection{\textbf{Domain-Adapted Parameters}}\label{subsubsec:domain_adpated_parameter}
To evaluate the impact of domain-adapted initialization, we conduct an ablation study within the Light-FMP framework across several backbone models. Figure~\ref{fig:init_ablation} shows the performance improvement (\%) in AUC and Logloss relative to standard initialization. While significant gains are observed for Wide\&Deep, and DNN, improvements remain marginal for xDeepFM, DeepFM and PNN, with PNN showing a slight 0.04\% performance degradation in LogLoss. 
The observed variation in performance gains from domain-adapted initialization across model architectures can be attributed to their differing reliance on initial feature representations versus learned interaction mechanisms. 
Models like DNN and Wide\&Deep depend more heavily on embedding quality and thus benefit substantially from domain-adapted initialization. In contrast, architectures with built-in interaction mechanisms, such as xDeepFM, DeepFM, and PNN, can compensate for suboptimal embeddings through explicit interaction learning, resulting in smaller gains. These findings confirm that domain-adapted initialization is most effective for representation-sensitive architectures, reinforcing its selective utility within the Light-FMP framework.

\begin{figure}[!t]
\centering
\includegraphics[width=0.48\textwidth]{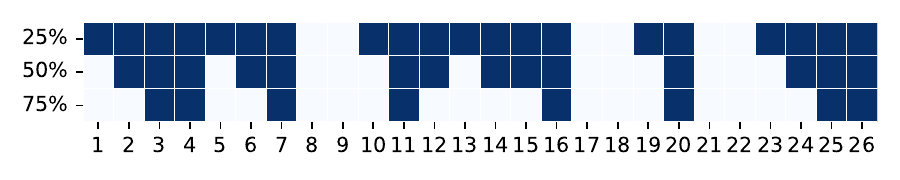} 
  \caption{Feature selection of Light-FMP with different prune ratios \( \tau = \{0\%, 25\%, 50\%, 75\%\} \) using xDeepFM on Criteo dataset. Color represents the normalized masking variable values of each feature field after pretraining, the darker the more important. The retained features are marked as blue color.}
\label{fig:heat_map}
\end{figure}

\subsection{Effectiveness of Masking Variable \texorpdfstring{$\mathbf{z}$}{Lg}} \label{subsec:effectiveness_z}
The masking variable plays a crucial role in our framework. To validate its effectiveness in prioritizing impactful features and ensuring that feature selection is driven by data relevance rather than arbitrary choices, we extract feature importance scores from pretraining. Figure~\ref{fig:heat_map} illustrates a heatmap of feature importance, determined by normalized \(\mathbf{z}\) values, across various pruning ratios using the xDeepFM model on the Criteo dataset.

From the figure, we observe that certain features (e.g., C3, C4, C7, C11, C16, C20, C25, C26) are consistently retained across varied pruning ratios, even when 75\% of features are pruned, indicating they are the most impactful in the input feature space in our experiment. Moderately important features (e.g., C2, C6, C12, C14, C15, C24) are pruned at higher ratios. Features like C8, C9, C17, C18, C21, and C22 are consistently pruned, suggesting they are the least important. 
This demonstrates a consistent pattern in how the masking variable is learned given a pretraining dataset and how it selects key features at different sparsity levels during the pruning phase. 


\section{Conclusion}

We propose Light-FMP, a lightweight framework that addresses the dual challenges of accuracy and efficiency in DRS. By designing super efficient pretraining and domain-adapted initialization, Light-FMP can reduces computational cost while improving predictive performance. 
Extensive experiments on benchmark datasets demonstrate that Light-FMP achieves a consistent trade-off between training efficiency and model accuracy across different tasks compared to existing shallow and deep baselines. Its modular design and broad applicability make Light-FMP a practical and scalable solution for real-world recommendation tasks.

A potential limitation of Light-FMP is that its efficiency improvement may be less significant for backbone models with large or deep hidden layers, as masking is currently applied only to the embedding layer. Extending masking to hidden layers for structured pruning could further reduce model complexity and bridge the gap between feature selection and model compression. 
However, this poses challenges in preserving predictive performance, as removing entire structures may disrupt key representations. We leave this exploration for future work.

\bibliographystyle{named}
\bibliography{ijcai26}

\end{document}